\documentclass[aps,prl,twocolumn,notitlepage,superscriptaddress,showpacs,10pt, nofootinbib]{revtex4-1} 
\usepackage{amsmath}    
\usepackage{mathtools} 
\usepackage{amsfonts}
\usepackage{bm}
\usepackage{amssymb}
\usepackage{appendix}
\usepackage{graphicx}   
\usepackage{xcolor}      
\usepackage{subfigure}  
\usepackage{comment}
\usepackage{siunitx}
\usepackage{cancel}

\usepackage{soul}
\usepackage[normalem]{ulem}

\mathchardef\mhyphen="2D

\begin{document}

\title{Reentrant delocalization transition in one-dimensional photonic quasicrystals}
\author{Sachin Vaidya}
\thanks{These authors contributed equally}
\affiliation{Department of Physics, The Pennsylvania State University, University Park, Pennsylvania 16802, USA}

\author{Christina J{\" o}rg}
\thanks{These authors contributed equally}
\affiliation{Department of Physics, The Pennsylvania State University, University Park, Pennsylvania 16802, USA}

\author{Kyle Linn}
 \affiliation{Department of Physics, The Pennsylvania State University, University Park, Pennsylvania 16802, USA}

\author{Megan Goh}
 \affiliation{Department of Physics, Amherst College,Amherst, MA 01002, USA}

\author{Mikael C. Rechtsman}
\affiliation{Department of Physics, The Pennsylvania State University, University Park, Pennsylvania 16802, USA}

\date{\today}

\begin{abstract}
Waves propagating in certain one-dimensional quasiperiodic lattices are known to exhibit a sharp localization transition. We theoretically predict and experimentally observe that the localization of light in one-dimensional photonic quasicrystals may be followed by a second delocalization transition for some states on increasing quasiperiodic modulation strength - an example of a reentrant transition. We further propose that this phenomenon can be qualitatively captured by a dimerized tight-binding model with long-range couplings.
\end{abstract}

\maketitle
Anderson localization is a generic phenomenon of wave localization in randomly disordered media \cite{Anderson_localization_original}. The presence of localized states implies the cessation of all wave transport in the thermodynamic limit and thus the Anderson model has provided deep insights into the nature of metal to insulator transitions for electrons in disordered solids \cite{fifty_years_of_anderson} as well as for light propagating in disordered photonic structures \cite{Anderson_localization_of_light}. Specifically in photonics, localization has been proposed and observed in photonic crystals (PhCs) and waveguide arrays, both in truly random \cite{john1987_PhC, Maret_Loc, genack_loc, lahini2008anderson_loc,Anderson_localization_of_light, Anderson_PhC_waveguide} and quasicrystalline cases \cite{AA_waveguides1, PhC_aubry_Andre1, PhC_aubry_Andre2}. Furthermore, this localization phenomenon can be employed for various photonic applications, such as for random nanolasing \cite{PhC_anderson1}, formation of photonic pseudogaps \cite{Anderson_PhC_bandgap}, formation of high $Q/V$ nanocavities \cite{PhC_anderson2, PhC_anderson3, PhC_aubry_Andre1, PhC_aubry_Andre2} and for reducing the crosstalk between waveguides in fiber arrays for endoscopy and telecommunications \cite{guglielmon2019inducing}.


It can be shown that in one and two dimensions, an infinitesimal amount of random disorder causes wave localization but in three dimensions, a sharp transition occurs between extended and localized regimes at a finite value of disorder strength \cite{Anderson_localization_1D, Anderson_localization_2D}. Such a sharp transition between localized and extended regimes can also occur in one dimension when the random disorder of lattice potentials is replaced by quasiperiodicity. A model first proposed by Aubry and Andr\'e consists of a one-dimensional lattice with quasiperiodic on-site-energy modulation and nearest-neighbor couplings that exhibits a sharp localization transition \cite{AA_original}. Specifically, the on-site potential for the $n$-th site in a chain of atoms is modulated according to $E_n = E_0 + \xi \cos(2\pi \beta n)$, where $E_0$ is the unperturbed on-site energy, $\beta$ is an irrational number and $\xi$ is the strength of the quasiperiodic modulation. For this simple model, the localization transition occurs for the entire spectrum at a single value of $\xi$ due to a duality between the extended and localized regimes \cite{AA_original}. 

Extensions of the Aubry-Andr\'e model with long-range couplings \cite{AA_longrange1, AA_longrange2, AA_generic_critical} and non-Hermiticity \cite{AA_NH1, reentrant4} were investigated theoretically and found to possess single-particle mobility edges and consequently intermediate regimes, where both extended and localized states co-exist. Moreover, some dimerized tight-binding models were recently found to exhibit a second reentrant transition of some states back to the same localization regime \cite{reentrant1, reentrant2, reentrant3, reentrant4}. Simple two-component one-dimensional PhCs can be thought of as naturally dimer-like due to the patterning of their different dielectrics and may exhibit non-Hermiticity from gain or radiative loss. They are therefore a potentially useful platform for exploring the rich localization physics in complex models.

In this work, we experimentally demonstrate a surprising localization phenomenon in multi-layer structures with quasiperiodic thickness modulation, i.e., one-dimensional photonic quasicrystals (PhQCs). In particular, we observe that in addition to the complete inhibition of transmission corresponding to a sharp localization transition, there is a second transition to an extended regime upon increasing the quasiperiodic modulation strength. The experimental signature of this is the complete recovery of transmission through the structure as the quasiperiodic modulation increases beyond the localized regime. This reentrant delocalization transition is not known to occur in random potentials and is a unique feature of quasicrystalline systems. To further explore the reentrant transition, we develop a tight-binding model inspired by our PhQCs, that captures the physics of localization and delocalization in our system.

\begin{figure}[t]
    \centering
    \includegraphics[width=\linewidth]{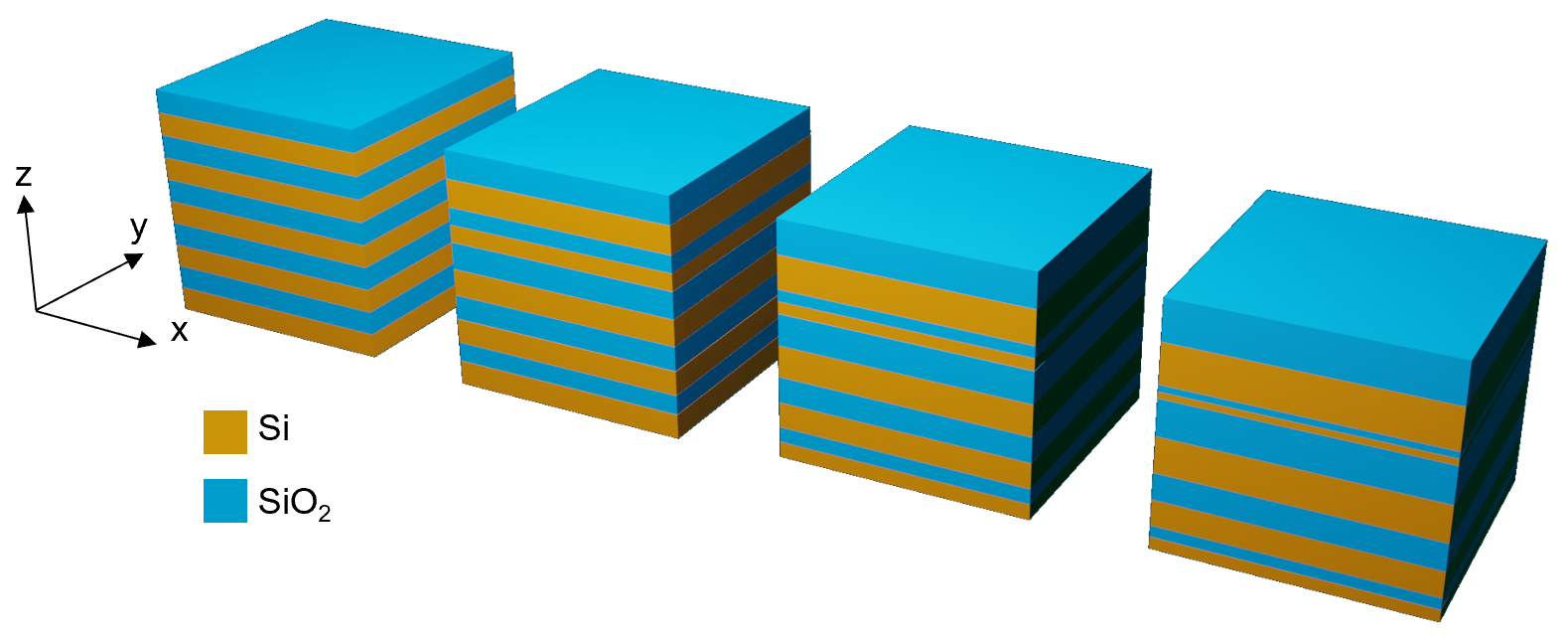}
    \caption{ \small{ Schematic of multi-layer photonic structures made out of Si and SiO$_2$ layers. The structures have increasing quasiperiodic modulation of layer thicknesses (from left to right); the leftmost structure is a perfect one-dimensional photonic crystal and the rest are photonic quasicrystals.
      \label{fig:1}}}
\end{figure}

The system considered here is shown in Fig. \ref{fig:1} and consists of a set of multi-layer structures made out of two materials, silicon and silica (SiO$_2$), with refractive indices $n_{\text{Si}}=3.5$ and $n_{\text{SiO}_2}=1.5$, respectively. These layers are stacked along the $z$-direction and define the dielectric function, $\epsilon(z)$. When propagation purely along the $z$-direction is considered, this system is described by the following Maxwell eigenvalue problem for a single scalar field $\mathcal{H}(z)$ \cite{photoniccrystalsbook, photoniccrystalsbook2}:
\begin{align}
-\partial_z\left( \frac{1}{\epsilon(z)}  \partial_z\right)\mathcal{H}(z) =&\left(\frac{\omega}{c}\right)^2 \mathcal{H}(z),
\label{eq:maxwell}
\end{align}
where $\mathbf{H}_{\text{TE}} = \mathcal{H}(z) \mathbf{\hat{x}}$ and $\mathbf{H}_{\text{TM}} = \mathcal{H}(z) \mathbf{\hat{y}}$ are the degenerate TE- and TM-polarized magnetic field solutions respectively, with frequency eigenvalue $\omega$.

Motivated by the Aubry-Andr\'e model, we modulate the thicknesses of each layer in a unit cell, defined as a pair of neighboring Si and SiO$_2$ layers, according to
\begin{align}
    t_n = t_0[1 + A\cos(2\pi \beta n)],
    \label{eq:thickness}
\end{align}
where $n \in \{1, 2, ..., N\}$ identifies a pair of layers, $2N$ is the total number of layers, $A$ is the strength of the spatial modulation and $\beta$ is the closest Diophantine (rational) approximation to the golden mean, $\phi = (1+\sqrt{5})/2$, for a given value of system size, $N$. 

When $A=0$, all layers have the same thickness $t_0$ and the system is a 1D PhC with a lattice constant of $a = 2t_0$, whereas for non-zero values of $A$, the integer sampling frequency of the cosine term and the irrational modulation frequency $\beta$ provide two competing and incommensurate periods that result in a 1D PhQC. In the latter case, the average lattice constant $\langle a \rangle = 2t_0$ provides a convenient length scale. We note that since $A$ modulates the thicknesses of layers, it is a bounded parameter with $|A| \le 1$.

\begin{figure}[]
    \centering
    \includegraphics[width=\linewidth]{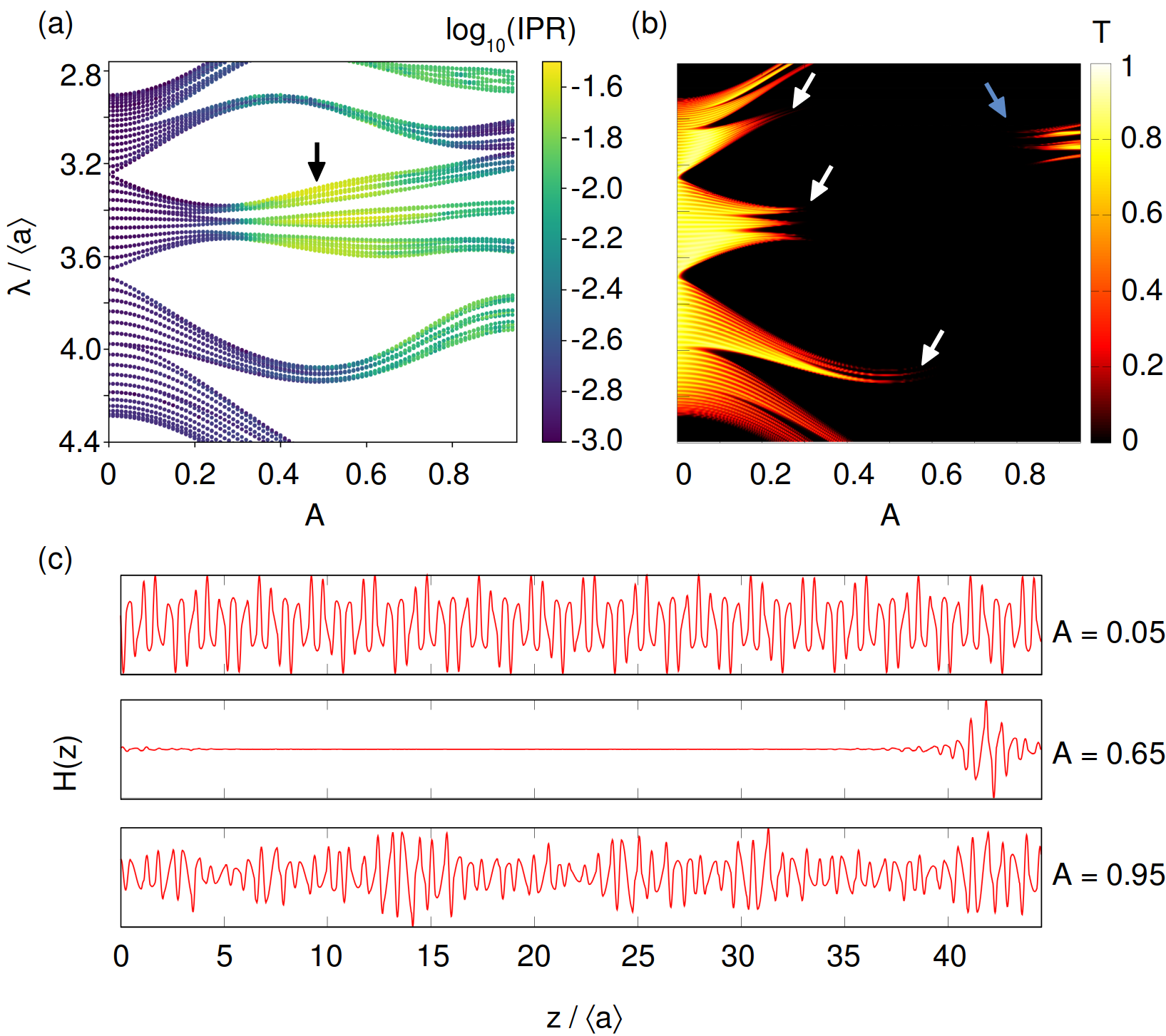}
    \caption{ \small{(a) Eigenvalue spectrum of the PhQC states and their corresponding IPR as a function of $A$ for $N = 89$. (b) The transmission spectrum as a function of $A$ for $N = 89$. Localization of various states corresponds to sharp drops in transmission (white arrows). Some states undergo a second delocalization transition around $A = 0.8$, which results in a sharp recovery of transmission (blue arrow). (c) $\mathcal{H}(z)$-field profiles of the state marked with the black arrow in (a), for various values of $A$.
      \label{fig:2}}}
\end{figure}

We obtain the states of our PhQCs using the plane-wave expansion method, as implemented in the open source software package MIT Photonic Bands (MPB) \cite{MPB}, and calculate their inverse participation ratios (IPR) given by
\begin{align}
\text{IPR}_p = \frac{\int|\mathcal{H}_p(z)|^4 dz}{\left[\int|\mathcal{H}_p(z)|^2 dz\right]^2},
\end{align}
where $\mathcal{H}_p$ is the scalar field in \eqref{eq:maxwell}, corresponding to the $p$-th state and the integral is taken over the entire finite system. IPR is a measure of localization of states, where small (large) values of IPR indicate extended (localized) states. 

The results for a system size of $N=89$ are shown in Fig. \ref{fig:2}. Figure \ref{fig:2} (a) shows a plot of the eigenvalue spectrum of the PhQC states as a function of $A$ and their corresponding IPR. In this plot, we focus on states corresponding to the second band in the PhC limit (i.e., at $A = 0$), and convert their corresponding frequency eigenvalues to dimensionless wavelength. For small values of $A$ ($A < 0.3$), the states are extended since the structure may be thought of as being crystalline with a small quasicrystalline perturbation. For larger values of $A$, the states undergo transitions to a localized regime, as indicated by a sharp increase in their IPR. However, as seen from Fig. \ref{fig:2} (a), these transitions do not all occur at the same value of $A$. Moreover, for some states around $\lambda/\langle a \rangle =  3.2$ and $A = 0.8$, we observe a sharp reduction in IPR on further increasing $A$, marking a reentrant transition to a second extended regime for these states. In Fig. \ref{fig:2} (c), we also examine the $\mathcal{H}(z)$-field profile for one such state that undergoes a reentrant transition, marked by the arrow in Fig. \ref{fig:2} (a). The field profiles show the transition from extended to localized and back to extended as $A$ is increased. 

Our system thus exhibits some crucial distinctions from the simple Aubry-Andr\'e model. Each pair of layers that forms a unit cell in our PhQCs is not well approximated as a resonator or atomic potential that is evanescently coupled only to its nearest neighbors. If the PhQC corresponds to a tight-binding model at all, it must be thought of as possessing long-range couplings that can be accurately computed using Wannier-function methods \cite{PhC_wannier1}. The presence of these effectively long-range couplings creates single-particle mobility edges that result in intermediate regimes where both extended and localized states co-exist \cite{AA_longrange1}. In fact, we find that due to the bounded nature of the quasiperiodic modulation strength via the parameter $A$, a large part of the spectrum of the PhQC is in an intermediate regime \cite{SupplMat, AA_generic_critical}. Moreover, the states corresponding to the lowest band of the PhC limit never localize for any value of $A$ up to its bounds \cite{SupplMat}. This is because PhQCs act as effectively homogeneous dielectric media at long wavelengths. Finally, the presence of a reentrant transition suggests the breakdown of the duality between the localized and extended regimes that exists in the simple Aubry-Andr\'e model.

Since the localization of states causes the cessation of wave transport, we explore its consequences in the transmission spectrum of the PhQCs. Fig. \ref{fig:2} (b) shows a plot of the transmission spectrum of the PhQCs as a function of $A$, calculated using a transfer matrix approach. We see that the localization transitions from Fig. \ref{fig:2} (a) correspond to the vanishing of transmission through the structures. This occurs because the system size is larger than the localization length of the localized states and as such these states are unable to form a transmission channel across the structures. Furthermore, we also see a recovery of transmission around $\lambda/\langle a \rangle \sim  3.2$, which corresponds to the reentrant transition of some of the states to an extended regime. This observation is not a finite size effect and persists for much larger system sizes \cite{SupplMat}. PhQCs and their transmission spectra therefore provide an accessible experimental setting in which to explore localization phenomena in one-dimensional systems.

\begin{figure}[t]
    \centering
    \includegraphics[width=\linewidth]{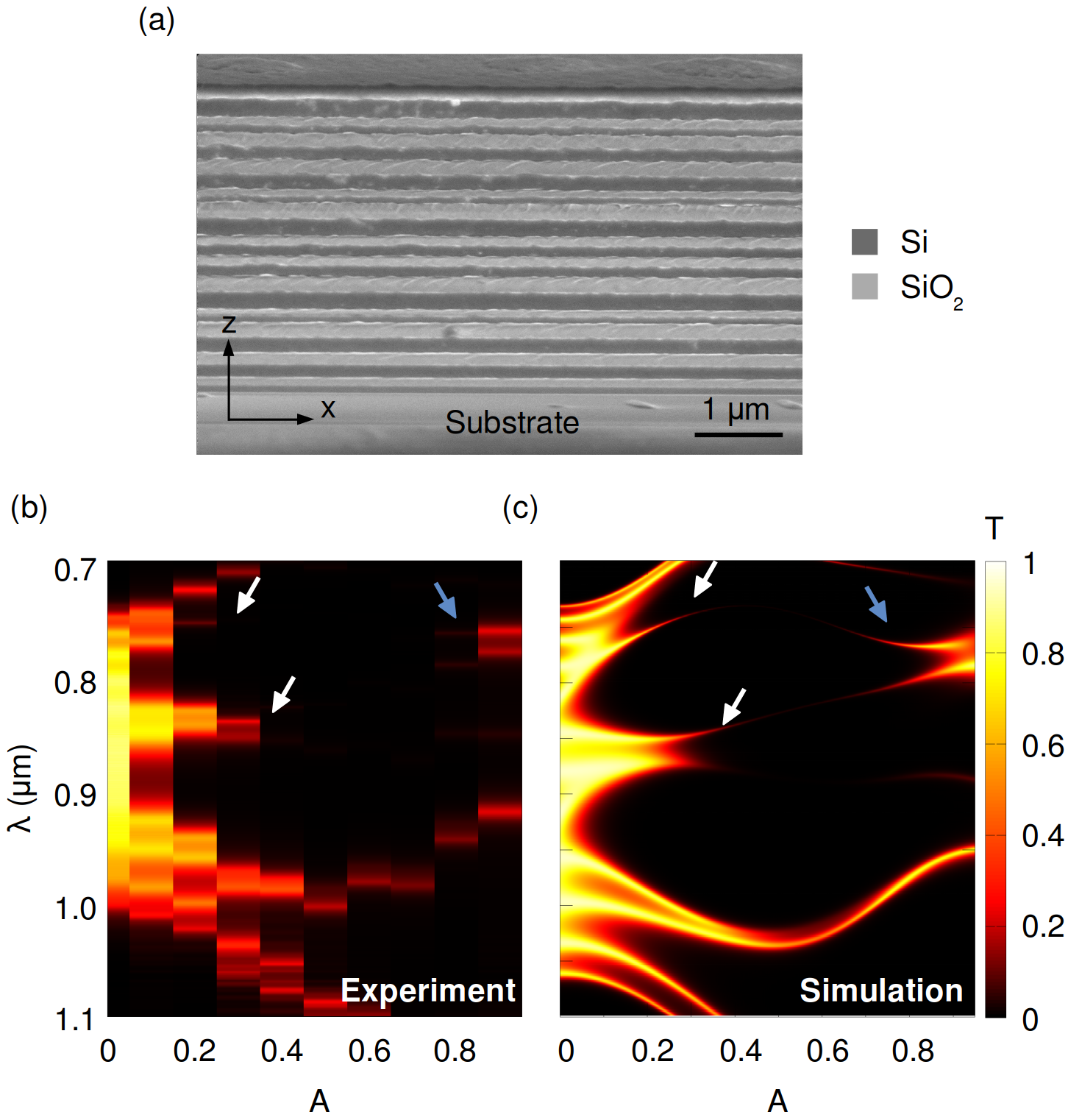}
    \caption{ \small{(a) Scanning electron microscope (SEM) image of a cut through a typical one-dimensional photonic quasicrystal fabricated by PECVD. Both layers in a pair of neighboring Si and SiO$_2$ layers have identical thicknesses. The thickness values of each such pair are modulated according to Eq. \eqref{eq:thickness}. (b) Experimentally measured transmission spectrum as a function of $A$ for $N = 13$. (c) Simulated transmission spectrum as a function of $A$ for $N = 13$. In (b) and (c), the localization transitions are marked with white arrows and the reentrant delocalization transition is marked with a blue arrow. \label{fig:3}}}\end{figure}

For the experiment, we fabricate the PhQCs using plasma-enhanced chemical vapor deposition (PECVD), alternating between Si and SiO$_2$ deposition on a glass substrate. The deposition times control the thicknesses of each layer and are determined from \eqref{eq:thickness}. We fabricate a total of ten samples with $\langle a \rangle =$ \SI{0.25}{\micro\meter}, $t_0 =$ \SI{0.125}{\micro\meter}, $N=13$, $\beta = 21/13$ and varying values of $A$. A scanning electron microscope (SEM) image of a typical sample is shown in Fig. \ref{fig:3} (a). To characterize our samples, we measure the transmission spectrum of each sample as a function of wavelength, normalized to the transmission through the bare glass substrate. This is performed using a supercontinuum laser in combination with a filter that allows for wavelength selection in the range of \SI{690}{\nano\meter} to \SI{1100}{\nano\meter}. The transmitted power is measured with a photodiode power sensor. 

The measured transmission spectrum is shown in Fig. \ref{fig:3} (b), along with the simulation results for comparison in Fig. \ref{fig:3} (c). We find that despite the relatively small system size, the localization transitions for states near $\lambda \sim$ \SI{0.85} and \SI{0.75} {\micro\meter} are clearly observed as the sharp inhibition of transmission. Furthermore, the second transition to an extended regime near $\lambda \sim$ \SI{0.78}{\micro\meter} is observed as a sharp recovery of transmission for $A>0.8$. The states near $\lambda \sim$ \SI{1}{\micro\meter} are localized for $A>0.5$, however the system size in the experiment is smaller than the localization length of these states and as such we measure finite transmission around this wavelength \cite{SupplMat}. Therefore it would be possible in principle to extract the localization length of states directly from the transmission spectrum of PhQCs by varying the system size.

\begin{figure}[t]
    \centering
    \includegraphics[width=\linewidth]{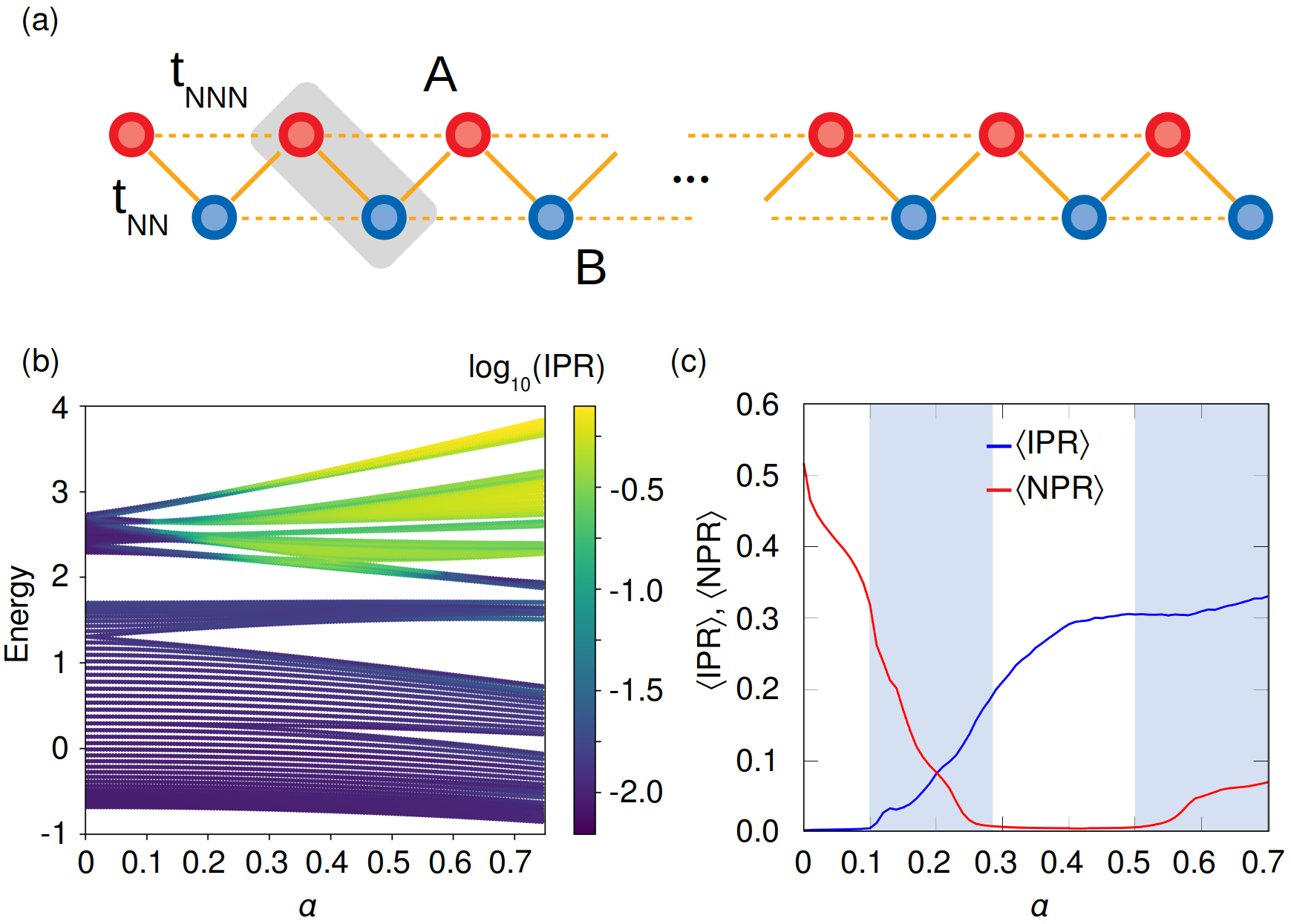}
    \caption{ \small{(a) Schematic of the tight binding model. The dimerized unit cell for the periodic system ($\alpha = 0$) is highlighted. The solid (dotted) lines represent nearest-neighbor (next-nearest-neighbor) couplings. (b) The energy spectrum and IPR of the corresponding states of the model for $E_{i,A} = 1$, $E_{i,B} = 2$, $t_{\text{NN}} = 0.7$, $t_{\text{NNN}} = 0.35$ and $N = 89$. The states of the second band exhibit a mobility edge and are localized for $0.25 < \alpha < 0.6$. Some states of this band undergo a reentrant delocalization transition at $\alpha \sim 0.6$ (c) A plot of the $\langle \text{IPR} \rangle$ and $\langle \text{NPR} \rangle$ for the states of the second band. The highlighted areas indicate intermediate regimes, where both $\langle \text{IPR} \rangle$ and $\langle \text{NPR} \rangle$ are non-zero, and localized and extended states co-exist. \label{fig:4}}}\end{figure}

To further explore the observed localization features, we develop a tight-binding model that qualitatively captures the physics of localization in our PhQCs. In particular, we consider a 1D quasiperiodic model with nearest and next-nearest neighbor couplings given by the Hamiltonian
\begin{multline}
H = \sum_{j = A, B } \sum_{i=1}^{N} E_{i,j} \left[ 1 + \alpha \cos(2\pi\beta i) \right]n_{i,j} \\ -t_{\text{NN}} \sum_{i=1}^{N} \left( c^{\dag}_{i,A}c_{i,B} + \text{h.c.} \right) -t_{\text{NN}} \sum_{i=1}^{N-1} \left( c^{\dag}_{i,B}c_{{i+1},A} + \text{h.c.} \right) \\ - t_{\text{NNN}} \sum_{j = A, B } \sum_{i=1}^{N-1} \left( c^{\dag}_{i,j}c_{i+1,j} + \text{h.c.} \right),
\end{multline}
where $c^{\dag}_{i,j}$, $c_{i,j}$ and $n_{i,j}$ are respectively the creation, annihilation and number operators on site $i$ of sublattice $j = A, B$. $t_{\text{NN}}$, $t_{\text{NNN}}$ are the nearest- and next-nearest-neighbor couplings respectively and $E_{i,j}$ are the unperturbed on-site energies of site $i$ of sublattice $j$. This lattice of $2N$ sites is shown schematically in Fig. \ref{fig:4} (a). We choose $E_{i,A} = 1$ as the energy scale and set $\beta = (1+\sqrt{5})/2$. $\alpha$ is an unbounded parameter that governs the strength of the quasiperiodic modulation of the on-site energies and we choose $E_{i,B} \ne E_{i,A}$ to introduce dimerization, akin to the two different layers in the PhQCs.

We plot the IPR of the states of this model for $N = 89$ in Fig. \ref{fig:4} (b), where we observe some important qualitative similarities with our PhQCs. The states of the lower band stay extended until a much larger value of $\alpha$, compared to the upper band, similar to the lowest two bands in the PhQCs. We also observe that the states of the upper band exhibit mobility edges and an intermediate regime, before undergoing a transition to a completely localized regime at $\alpha \sim 0.25$. Moreover, some states of this band undergo a second transition at $\alpha \sim 0.6$ and remain extended for a range of $\alpha$ values. Eventually all states of this model become localized for a large enough value of $\alpha$ ($\alpha > 2$). We find that these features are generic and persist for a range of parameters of the model.

We also calculate the average IPR and average normalized participation ratio (NPR) for a set of $M$ states, given by
\begin{align}
    \langle \text{IPR} \rangle =& \frac{1}{M} \sum_{n = 1}^M \sum_{i=1}^{2N} |\psi_{n,i}|^4, \\ \langle \text{NPR} \rangle =& \frac{1}{M} \sum_{n = 1}^M  \left(2N \sum_{i=1}^{2N} |\psi_{n,i}|^4 \right)^{-1},
\end{align}
where $|\psi_{n,i}\rangle$ is the normalized $n$-th eigenstate of $H$ and $i$ labels the sites. The extended regime is characterized by near-zero $\langle \text{IPR} \rangle$ and non-zero $\langle \text{NPR} \rangle$ and vice versa for the localized regime. A non-zero value for both $\langle \text{IPR} \rangle$ and $\langle \text{NPR} \rangle$ implies the presence of an intermediate regime in the spectrum. The plot of $\langle \text{IPR} \rangle$ and $\langle \text{NPR} \rangle$ for the states of the upper band is shown in Fig. \ref{fig:4} (c). Comparing this plot with Fig. \ref{fig:4} (b), we can see that the first intermediate regime arises due to a moblity edge and the second intermediate regime arises due to a reentrant delocalization transition for the lowest lying states of the band. 

Through this model, we see that a combination of staggered potentials and long-range couplings, competing with quasiperiodicity, can cause the delocalization of previously localized states for a range of parameter values. These findings are consistent with other models with similar qualities that are known to host reentrant transitions \cite{reentrant1, reentrant3}.

In conclusion, we have observed a reentrant delocalization transition -- a feature that is not present in the standard Aubry-Andr\'e model -- in 1D PhQCs with an Aubry-Andr\'e-type quasiperiodic modulation by measuring their transmission spectra. The PhQCs and their transmission spectra thus provide a means to experimentally explore more complex models with richer localization physics, as compared with simple nearest-neighbor tight-binding models. Inspired by the PhQCs, we have also explored the localization features of our system in a tight-binding setting, in order to lend physical insight into the nature of the transition. In the future, it will be interesting to explore localization in passive non-Hermitian 1D PhQCs enabled via a patterning of lossy dielectric materials. Furthermore, examining localized states in higher dimensional realizations of PhQCs is warranted since it could lead to better-performing photonic nanocavities with lower index materials.

We thank Zeyu Zhang for SEM images of the PhQCs and Bill Mahoney for technical help with the PECVD process.
C.J. acknowledges funding from the Alexander von Humboldt Foundation within the Feodor-Lynen Fellowship program. 
M.C.R. and S.V. acknowledge the support of the Charles E. Kaufman Foundation under Grant No. KA2020-114794, the U.S. Office of Naval Research Multidisciplinary University Research Initiative (MURI) under Grant No. N00014-20-1-2325 as well as the U.S. Army Research Office MURI under Grant No. W911NF-22-2-0103. K. L. and M. G. acknowledge support from the Pennsylvania State University REU program.

\bibliography{references.bib}    


\end{document}


\title{Supplemental Material for: Reentrant delocalization transition in one-dimensional photonic quasicrystals}
\author{Sachin Vaidya}
\affiliation{contributed equally}
\affiliation{Department of Physics, The Pennsylvania State University, University Park, Pennsylvania 16802, USA}

\author{Christina J{\" o}rg}
\affiliation{contributed equally}
\affiliation{Department of Physics, The Pennsylvania State University, University Park, Pennsylvania 16802, USA}

\author{Kyle Linn}
 \affiliation{Department of Physics, The Pennsylvania State University, University Park, Pennsylvania 16802, USA}

\author{Megan Goh}
 \affiliation{Department of Physics, Amherst College,Amherst, MA 01002, USA}

\author{Mikael C. Rechtsman}
\affiliation{Department of Physics, The Pennsylvania State University, University Park, Pennsylvania 16802, USA}

\date{\today}
\maketitle

\section{Larger system size and finite size effects}

\begin{figure}[h]
    \centering
    \includegraphics[width=0.75\linewidth]{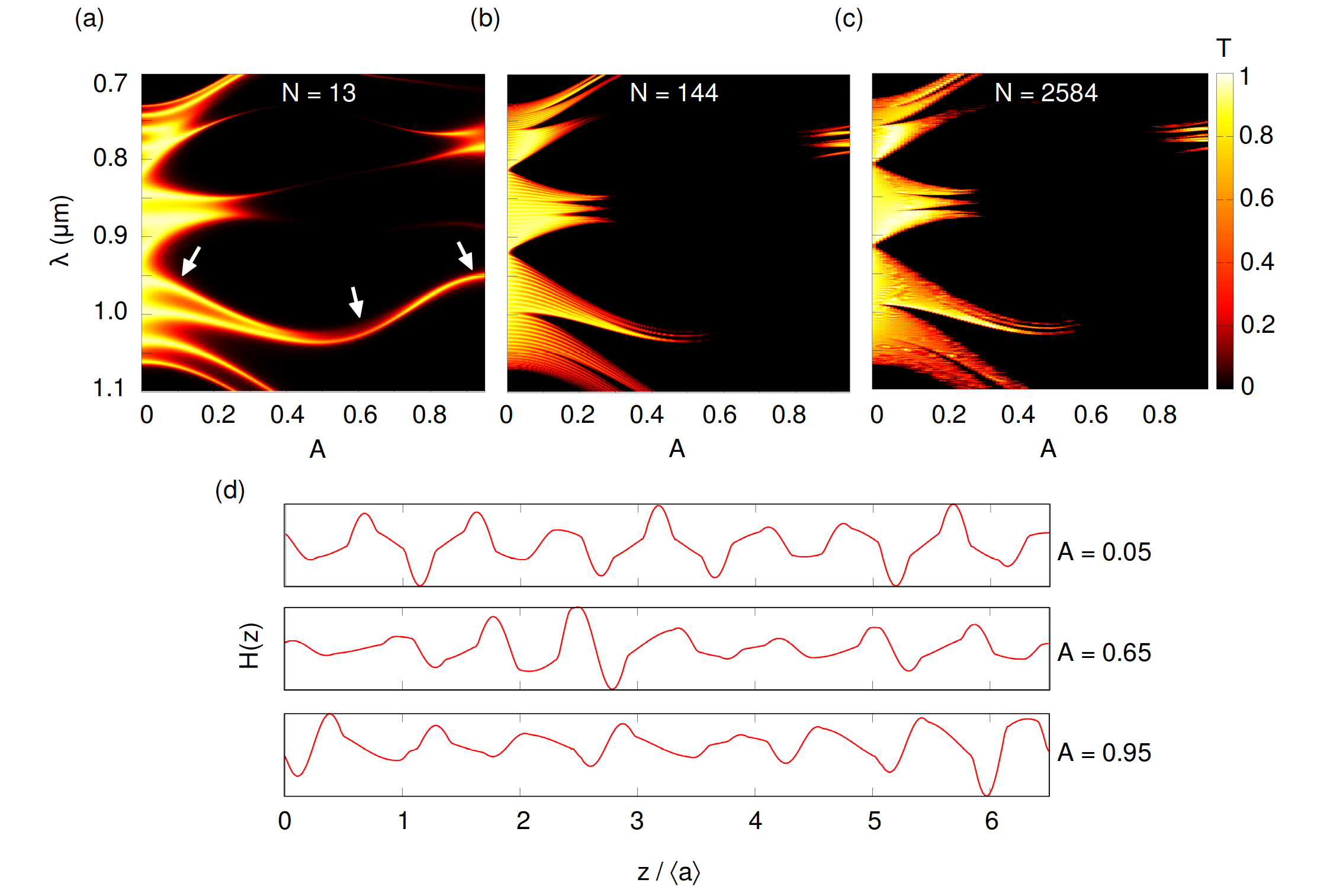}
    \caption{ \small{(a) Transmission spectrum for $N=13$. This is the system size fabricated in the experiment. The white arrows mark a transmission channel that corresponds to a weakly localized state whose localization length is comparable to the system size. (b) Transmission spectrum for $N= 144$ (c) Transmission spectrum for $N=2584$. The reentrant transition occurs for arbitrarily large system sizes. (d) The $\mathcal{H}(z)$-field profile of the state that corresponds to the transmission channel marked with white arrows in (a), for various values of $A$.
      \label{fig:S1}}}
\end{figure}

In the main text, we mention that the reentrant transition is not a small-system-size effect but persists for much larger system sizes. Here we show this using the simulated transmission spectra for a very large system size. Fig. \ref{fig:S1} (a), (b) and (c) show the transmission spectra for a system size of $N = 13$, $N = 144$ and $N = 2584$ respectively. In all three cases, we can see the inhibition of transmission associated with the localization transitions at $\lambda \sim$ \SI{0.85} and \SI{0.75} {\micro\meter}, $A \sim 0.3$ and the recovery of transmission associated with the reentrant delocalization transition at $\lambda \sim$ \SI{0.78}{\micro\meter}, $A \sim 0.8$. We note that all features in the spectrum are already well converged for $N = 144$.

At relatively small system sizes, such as the system size in the experiment ($N=13$), a transmission channel appears at $\lambda \sim$ \SI{1}{\micro\meter} that corresponds to a localized state for $A> 0.5$ as marked in Fig. \ref{fig:S1} (a). We find that this is due to the localization length of this state being comparable to the system size. In Fig. \ref{fig:S1} (d) we plot the field profile of the state that corresponds to this transmission channel for various values of $A$ and find that it is only weakly localized and is therefore able to transmit power across the length of the samples for any value of $A$.

\section{Intermediate regimes}
\begin{figure}[h]
    \centering
    \includegraphics[width=0.45\linewidth]{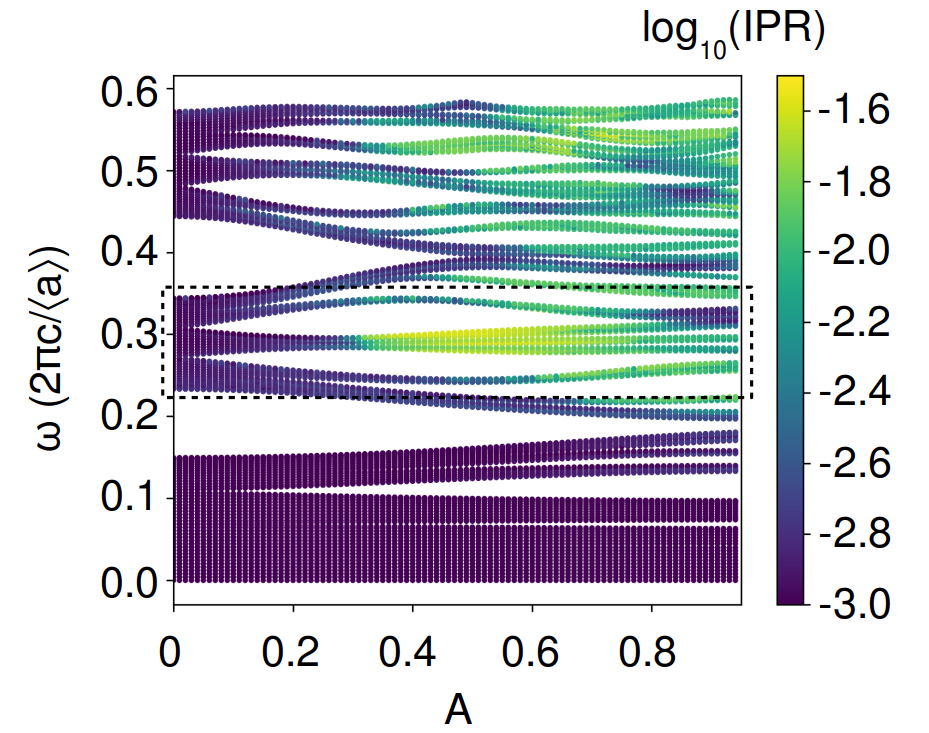}
    \caption{ \small{The eigenvalue spectrum for the family of PhQCs considered in the main text for a larger frequency range. In the main text and in the experiment, we focus on the localization features of states in the dimensionless frequency range of $0.227$-$0.357$ (marked by the box) that correspond to the second band of the PhC limit (i.e., at $A = 0$).
      \label{fig:S2}}}
\end{figure}

In the main text we state that much of the eigenvalue spectrum of the PhQCs is in an intermediate regime. In Fig. \ref{fig:S2}, we plot the spectrum for a larger frequency range along with the IPR of the corresponding states. We notice the presence of multiple mobility edges as well as multiple reentrant delocalization transitions for various values of $A$. Furthermore, we also see that the lowest band does not localize for the full range of $A$ due to the behavior of these PhQCs as effectively homogeneous dielectrics at long wavelengths.

\section{Methods}
For the fabrication of 1D PhQCs, we employ the plasma-enhanced chemical vapor deposition (PECVD) process to deposit alternating layers of silicon (Si) and silica (SiO$_2$). The layers are deposited onto a glass substrate (Corning \SI{18}{mm} square microscope glass cover slide). Si is deposited from Ar and SiH$_4$ precursor gases at \SI{220}{\degree{C}} and a pressure of 4.5 Torr, while silica is deposited from N$_2$O and SiH$_4$ precursor gases at \SI{300}{\degree{C}} and pressure of  3.5 Torr. The thicknesses of the layers are controlled by the deposition time. 

To characterize the fabrication imperfections in our system and show that the observed features are robust against fabrication disorder, we extract the layer thicknesses of one of the samples from SEM images and compare them with the targeted thicknesses given by Eq. (2) in the main text. We observe random fluctuations in the layer thicknesses with respect to the target thickness by a maximum of $\pm 8\%$ and an average of 2\%. This is most likely caused by the fabrication process, where the chemical reaction, and thus the layer formation, is controlled purely by a timed precursor release into the chamber (a process which inherently is prone to fluctuations). We find that these fabrication errors are not large enough to cause any meaningful deviation of the observed localization features compared to simulations.

For the measurements, a collimated, unpolarized laser beam is sent through the PhQCs at normal incidence, and the transmitted power is measured via a powermeter (Thorlabs S120c). To sweep through the wavelengths in the range of \SI{690}{nm} to \SI{1100}{nm}, a SuperK EVO white light laser (NKT Photonics) and a SuperK Select filter box are used. The transmitted power  is normalized to that from the bare glass substrate.

\begin{figure}[h]
    \centering
    \includegraphics[width=0.7\linewidth]{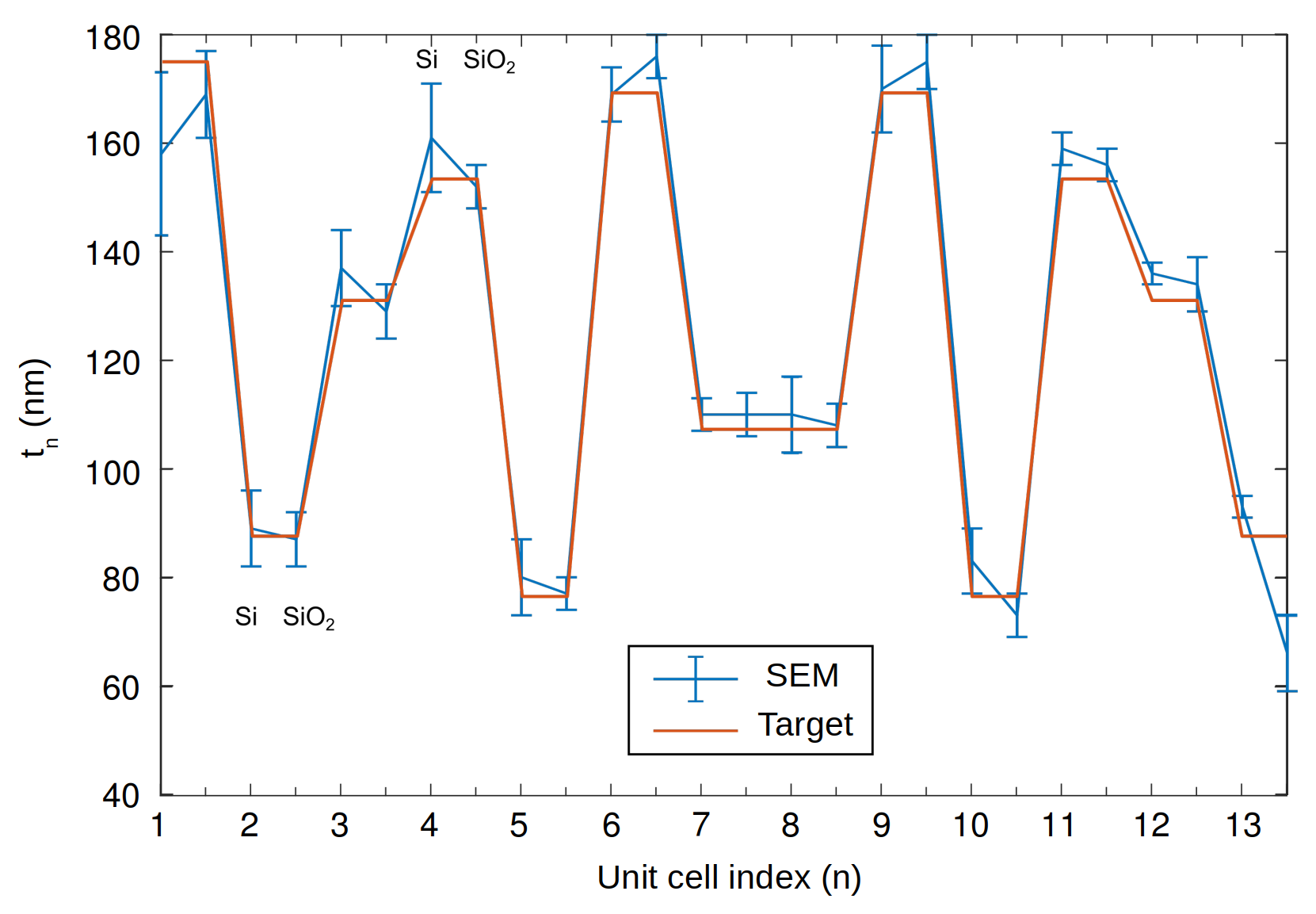}
    \caption{ \small{Comparison of a sample's layer thicknesses extracted from SEM images (blue) and targeted thickness (red).
      \label{fig:S3}}}
\end{figure}